\documentclass[a4paper,12pt]{article}

\usepackage{graphicx}
\input{epsf}

\newcommand{\beq}{\begin{eqnarray}}
\newcommand{\eeq}{\end{eqnarray}}

\newcommand{\ket}{\rangle}
\newcommand{\bra}{\langle}
\newcommand{\del}{\partial}

\newcommand{\dslash}{{\del \hspace{-5pt}/}}

\newcommand{\etal}{{\it et\ al.\ }}

\def\gsim{\displaystyle\mathop{>}_{\sim}}
\def\lsim{\displaystyle\mathop{<}_{\sim}}

\begin{document}

\begin{center}
{\large  \bf
Structure and reactions of pentaquark baryons}
 
 \vspace*{1cm}
 Atsushi Hosaka\\
 \vspace*{2mm}
 {\it Research Center for Nuclear Physics (RCNP), Osaka University\\
 Ibaraki 567-0047 Japan}
 %\vspace*{1cm}
 %\today
\end{center}
\vspace*{0.2cm}

\abstract{
    We review the current status of the exotic pentaquark 
	baryons.  
	After a brief look at experiments of both positive and 
	negative results, we discuss theoretical methods to study the 
	structure and reactions for the pentaquarks.  
	First we introduce the quark model and the chiral soliton model, 
	where we discuss the relation of mass spectrum and parity with 
	some emphasis on the role of chiral symmetry.  
	It is always useful to picture the 
	structure of the pentaquarks in terms of quarks.  
	As for other methods, we discuss a model independent method, and 
	briefly mention the results from the lattice and QCD sum rule.  
	Decay properties are then studied in some detail, 
	which is one of the important properties of $\Theta^+$.  
	We investigate the relation between 
	the decay width and the quark structure having 
	certain spin-parity quantum numbers.  
	Through these analyses, we consider as plausible quantum numbers 
	of $\Theta^+$, $J^P = 3/2^-$.  
	In the last part of this note, we discuss  
	production reactions of $\Theta^+$ which provide links 
	between the theoretical models and experimental information.  
	We discuss photoproductions and hadron-induced reactions 
	which are useful to explore the nature of $\Theta^+$. }

%===============================================
\section{Introduction}
%===============================================

History of exotic hadrons is as old as that of 
the quark model \cite{Gell-Mann:1964nj}, and
the subject has been studied for 
long time \cite{lipkin}.   
Yet the recent observation of 
the evidence of the pentaquark particle 
$\Theta^+$ has triggered 
enormous amount of research activities  both in experimental and 
theoretical hadron physics \cite{Diakonov:1997mm,Nakano:2003qx,penta04}.  
Baryons containing five valence quarks are totally new form 
of hadrons.   
The importance of knowing the nature of 
multi-quark states lies, for instance, in understanding
the origin of matter.  
It is believed that in the early stage of the universe,  
matter was highly dense forming the quark matter.
A natural question would then be what is the mechanism of the 
transition from that to the present hadronic world consisting 
of ordinary mesons and baryons.  

At this moment, the existence of the pentaquarks is still
the most important issue.  
The whole discussions below are, therefore, based on this 
assumption.  

From the hadron physics point of view, the understanding of five 
quark systems, if they exist as (quasi-)stable states, will 
give us more information on the dynamics of 
non-perturbative QCD, such as confinement of colors and 
chiral symmetry breaking.  
Many ideas have been proposed attempting to explain 
the unique features of $\Theta^+$.  
As it has turned out and will be discussed in this note, however, 
the current theoretical situation is not yet settled 
at all, having revealed that 
our understanding of hadron physics would be much poorer than 
we have thought \cite{penta04}. 
We definitely need more solid ideas and methods to answer 
the related questions.  

Turning to the specific interest in $\Theta^{+}$, 
its would-be light mass and narrow width
are the issues to be understood, 
together with the determination of its spin and parity.  
In particular, the information of parity is important, since it 
reflects the internal motion of the constituents.   

In this lecture the following materials are discussed, with 
emphasis on theoretical methods.  
\begin{itemize}
    \item  A quick overview on experimental situation (section 2).

    \item  Some basics of theoretical models; the quark model,  
	chiral soliton model \cite{Diakonov:1997mm}, 
	and somewhat general treatment based on the 
    flavor SU(3) symmetry as well as a brief view over lattice and 
	QCD sum rule studies (section 3).

    \item  Decay of $\Theta^{+}$ (section 4)
   
    \item Production reactions including photo and hadronic 
    productions (section 5).
\end{itemize}
Through these discussions, we consider a possibility of 
$\Theta^+$ with $J^P = 3/2^-$ as one of likely candidates
for a pentaquark state.  

There are many interesting topics which can not be discussed 
in this note.  
For readers who are interested in more details, 
please refer to the proceedings of the 
workshop PENTAQUARK04 and references in 
there \cite{penta04}.

%===============================================
\section{Experiments}
%===============================================

%-----------Figure--------------------
\begin{figure}[tbp] 
\epsfxsize=7cm
\centerline{\epsfbox{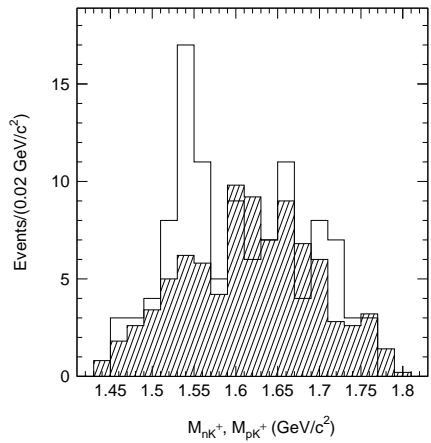}}
\centering
\caption{Invariant mass spectrum of the $nK^+$ 
extracted from the missing mass analysis of 
$\gamma n \to K^- K^+ n$ (unhatched histogram) 
and that of  $pK^+$ 
extracted from the analysis of 
$\gamma p \to K^- K^+ p$ (hatched histogram)
\cite{Nakano:2003qx}.  
}
\label{leps}
\end{figure}
%-----------Figure--------------------

The first observation was made by the LEPS group at 
SPring-8 lead by T. Nakano \cite{Nakano:2003qx}.  
The backward compton-scattered photon of energy 2.4 GeV
produced at SPring-8 was used to hit a neutron target inside 
a carbon nucleus to produce a strangeness and antistrangeness 
pair ($K^{+}$ and $K^{-}$).  
The Fermi motion corrections were carefully analyzed, and 
then a missing mass analysis was performed for the $K^{+}n$
final state.  
They have seen an excess in the $K^{+}n$ invariant mass 
spectrum over the background at 4.6$\sigma$ level in the 
energy region 1.54 GeV.  
The width of the peak was as narrow as or less than the 
experimental resolution ($\sim 25$ MeV).  
The peak was then identified with the exotic pentaquark 
state of strangeness $S = +1$.  
The absence of the similar peak structure in the 
$K^{+}p$ system suggests the isospin of the state 
is likely to be $I = 0$.  

The spectrum of the LEPS experiment is shown in 
Fig. \ref{leps}, where the peak around 1.54 GeV is the 
first signal of the exotic particle \cite{Nakano:2003qx}.  
Immediately after the announce of this results, 
many positive signals follow 
\cite{Barmin:2003vv,Stepanyan:2003qr,Kubarovsky:2003fi,Barth:2003ja,Airapetian:2003ri}.  
Among them, the existence of another exotic baryon 
of strangeness $-1$,  $\Xi^{0, -, --}$, was also 
reported \cite{Alt:2003vb}.  
Major results of experiments so far 
are summarized in Table \ref{experiments}.  
From there, one can recognize that there is fluctuation 
in absolute values of the mass of $\Theta^+$, from 
1520 to 1550 MeV.  
It is often said that 
the fluctuation of order 30 MeV is large; 
it is about 30 \% level if measured from the $KN$ threshold.  

%---------------------------------------------
\begin{table}[tbp]
\centering
\caption{\label{experiments} \small Brief summary of the previous experiments}
%\vspace*{0.5cm}
{\small 
\begin{tabular}{ l l c l l}
\hline
Experiment & Reaction & Energy (GeV) & Mass (MeV) & Width (MeV)\\
\hline
LEPS & $\gamma n$ & $E_\gamma \sim 2.4$ & 
          $\Theta^+ \sim 1540$ & $\Gamma \lsim 25$ \\
     & $\gamma d$ &  & being analyzed
           &   \\
Graal & $\gamma p \to \eta p$ & $E_\gamma \lsim 1.5$ & 
          $N^*_5 \sim 1715$ &   \\
      &   &   & 
          $\Theta^+ \sim 1531$ &   \\
CLAS & $\gamma p$ & $1.6 \lsim E_\gamma \lsim 2.3$ & 
          $\Theta^+ \sim 1555 \pm 10$ &   $\Gamma \lsim 20$\\
     & $\gamma d$ & $1.5 \lsim E_\gamma \lsim 3.1$ & 
          $\Theta^+ \sim 1542 \pm 5$ &   \\
CLAS(g11) & $\gamma p$ & $1.6 \lsim E_\gamma \lsim 3.8$ GeV& --
           &   \\
          & $\gamma d$ &  & being analyzed
           &   \\
HallA & $ep$ & $E_\gamma \lsim 5$ & being analyzed &   \\
HERMES & $ed$ & $\sqrt{s} \sim 10 $ & 
          $\Theta^+ \sim 1527 \pm 2.3$ &  
		         $\Gamma = 17 \pm 9 \pm 3$\\
ZEUS   & $ep$ & $300 \lsim \sqrt{s} \lsim 319$ & 
          $\Theta^+ \sim 1521.5 \pm 1.5$ &  
		         $\Gamma = 6.1 \pm 1.6^{+2}_{-1.4}$\\
H1     & $ep$ & $300 \lsim \sqrt{s} \lsim 319$ & 
          $\Theta^+_C \sim 3100 $ &  \\
NA49   & $pp$ & $\sqrt{s} \sim 17.2 $ & 
          $\Xi^{--} \sim 1680 $ &  \\
       &      &   & 
          $\Theta^{+} = 1526 \pm 2 $ &  \\
COSY-TOF & $pp$ & $p \sim 2.95 $ & 
          $\Theta^+ =1530 \pm 5 $ &  $\Gamma \leq 18 \pm 4$ \\
E522(KEK) & $\pi^- p$ & $p \sim 1.95 $ & 
          $\Theta^+ =1530 \pm 5 $ &  \\
 & & & & \\
E690(Fermilab) & $pp$ & $\sqrt{s} = 800$ & 
          -- &  \\
CDF(Fermilab) & $\bar p p$ & $\sqrt{s} = 1960$ & 
          -- &  \\
Babar & $e^+ e^-$ &  & 
          -- &  \\
Belle & $e^+ e^-$ & $e^+(3.5) e^-(8)$ & 
          -- &  \\
\hline
\end{tabular}
}
\end{table}
%---------------------

After many positive signals were reported, 
negative results followed also, mostly from the analysis of 
high energy experiments \cite{bib:Pq:eprint,mjwang,mizuk}.  
These are also summarized in Table \ref{experiments}.  
At this moment there is not a theory consistently explain 
these data.  
The high energy experiments have much higher 
statistics than the low energy experiments and should be 
taken seriously.  
If $\Theta^+$ exists and can be seen only in the 
low energy (mostly in photoproductions) experiments, 
one needs to understand the production mechanism \cite{Karliner:2004gr}.  
A model for the suppression at high energies was proposed 
by Titov \etal \cite{Titov:2004wt}.  
If it does not, we also need to understand 
what the signals in the low energy experiments are for.  

Very recently, CLAS (g11) reported the 
null result in the reaction 
$\gamma  p \to \bar K^0 K^+ n$ \cite{DeVita:2005CLAS}.  
This has much larger statistics than the 
previous experiment performed at SAPHIA
by about factor twenty \cite{Barth:2003ja}.  
They extracted an upper limit of the 
$\Theta^+$ production cross section, 
$\sigma \lsim$ 1 -- 4 nb.  
The results, however, does not immediately 
lead to the absence of $\Theta^+$, since there 
could be a large asymmetry between the reactions 
from the proton and neutron \cite{Nam:2005uq,Nam:2005jz}.  
In general, photoproductions are large
for charge-exchange reactions, but the 
reaction $\gamma  p \to \bar K^0 K^ n$ is not the like.  
Experimental studies from the neutron with higher 
statistics is therefore very important.

%===============================================
\section{Theoretical methods}
%===============================================

In this section, we discuss the structure of 
the pentaquarks, especially of $\Theta^+$.  
Although the pioneering work of Diakonov {\it et. al.}
was performed in the chiral soliton model 
\cite{Diakonov:1997mm}, it is always 
instructive and intuitively understandable to work 
in the quark model \cite{okaptp,hosaka_book}.  
After a brief look at the basics of pentaquark 
structure in the quark model, we discuss some essences 
of the chiral soliton model.  
Results of the chiral solitons are then interpreted in terms of 
a quark model with chiral symmetry (the chiral bag model)
\cite{Hosaka:2003jv}.  

After the introduction of the two models, we  
discuss a model independent method based on flavor SU(3) 
symmetry, where possible spin and parity of 
$\Theta^+$ are investigated \cite{Pakvasa:2004pg,Hyodo:2005wa}.  
In the last two subsections, we briefly 
look at the lattice QCD and QCD sum rule.

%------------------------------
\subsection{Constituent quark model}
%------------------------------

This model has been successfully applied to the 
description of the conventional mesons and baryons for 
their masses and various transition amplitudes \cite{quark}.  
In this model,  a 
confining potential for quarks is introduced, 
which is usually taken to be a harmonic 
oscillator one, to prepare basis states as 
single particle states which valence quarks occupy.     
Then quark-quark interactions such as the 
spin-color interaction of one gluon exchange and 
the spin-flavor one of one-meson (the Nambu-Goldstone boson) 
exchange
are introduced as residual interactions.  
The interaction hamiltonian is then treated either 
perturbatively or diagonalized within a given model space.  
The role of various interactions for $\Theta^+$ 
has been investigated in the 
literatures \cite{okaptp,jennings,Shinozaki:2004bp,Takeuchi:2004xm}.  
Here, to make discussions simple, we consider 
what the structure of the five-quark 
states are like in a confining potential.  
The single particle states of the harmonic oscillator 
potential are denoted 
by the principal and angular 
momentum quantum numbers $(n,l)$.  
Using the spectroscopic notation, 
we express them as $0s, 0p, 1s$ and so on.  

Due to the many degrees of freedom of color (3), 
flavor (3) and spin (2), five quarks including one 
anti-quark ($\bar s$) can occupy the lowest ground 
state simultaneously.  
Thus, we denote the ground state of the five quarks as 
$(0s)^{5}$.  
If one quark is excited to a $p$-orbit, $(0s)^{4}0p$ 
and so on.  
The parity of the ground state is negative, since the 
antiquark carries negative parity, while the 
parity of the first excited state is positive.  

\begin{figure}[tbp] 
\epsfxsize=6cm
\centerline{\epsfbox{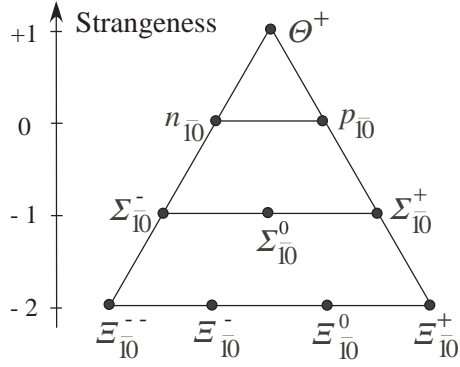}}
\centering
\caption{The weight diagram for $\bar{10}$. 
Particles belonging to this multiplet are also indicated. }
\label{antidec}
\end{figure}

Now, let us consider flavor structure.  
Under the assumption of SU(3) symmetry, we need to perform 
irreducible decomposition of the 
five quark states, the direct product of 
four fundamental and one conjugate representations, 
\beq
3 \otimes 3 \otimes 
3 \otimes 3 \otimes \bar 3 
=
1 \oplus 8 \oplus 10 \oplus \bar{10} \oplus 27 
\oplus 35\, , 
\label{decomp3333}
\eeq
where multiplicities are ignored on the right hand side.  
Among these representations on the right hand side, the 
isosinglet state of $S=+1$ appears only in the 
antidecuplet representation $\bar {10}$, which is  
a candidate for the SU(3) multiplet for the pentaquarks.  
The weight diagram of the  $\bar {10}$ representation is 
shown in Fig.~\ref{antidec}, where locations of 
various states are also indicated.  

%-----------Figure--------------------
%-----------Figure--------------------

Flavor wave functions of the antidecuplet states are 
easily constructed, if one notices that one of the five 
quark is $\bar 3$.  
Form two $\bar 3$'s out of two quark pairs (diquarks) 
in an antisymmetric combination, 
\beq
\bar Q_i = \epsilon_{ijk}q_j q_k \, ;  \; \; \; 
\bar U \sim [ds]\, , \; 
\bar D \sim [su]\, , \; 
\bar S \sim [ud]\, . 
\eeq
Then we can make symmetric products in terms of two diquarks and 
one antiquark, which generate antidecuplet members:
\beq
& & \Theta^+ = \bar S \bar S \bar s\, , \; \; \; 
N_{\bar{10}} = \frac{1}{\sqrt{3}} 
( \bar S \bar S \bar u + \bar S \bar U \bar s + \bar U \bar S \bar s) \, , 
\nonumber \\
& & \Sigma^-_{\bar{10}} = \frac{1}{\sqrt{3}} 
( \bar S \bar U \bar u + \bar U \bar S \bar u + \bar U \bar U \bar s) \, , 
\; \; \; 
\Xi^{--}_{\bar{10}} = \bar U \bar U \bar u\, .
\label{ad_members}
\eeq
There are analogous to the decuplet wave functions for 
($\Delta, \Sigma^*, \Xi^*, \Omega$).  

What is interesting is the average number of strange
(and anti-strange) quarks in the wave functions.  
One can easily verify that 
it is 1 for $\Theta^+$, 4/3 for $N_{\bar{10}}$, 5/3 for $\Sigma_{\bar{10}}$
and 2 for $\Xi_{\bar{10}}$.
Namely, the strange quark content increases by equal amount 
1/3 as the hypercharge decreases.  
This is a general consequence valid to  the symmetric 
representation of SU(3).  
The simple counting implies that if 
the $\lambda_8$ is the only source 
of the SU(3) breaking as  
$m_u = m_d << m_s$, where $m_i$ are constituent quark masses, 
the equal mass splitting of the antidecuplet baryons 
is expected to be $\sim (1/3) (m_s - m_u) \equiv (1/3) \Delta$.  
Hence we also expect 
\beq
M(\Xi_{\bar{10}}) - M(\Theta^+) \sim m_s - m_u \sim 200 \; {\rm MeV}\, .  
\label{massdiff}
\eeq

\begin{figure}[tbp] 
\epsfxsize=8cm
\centerline{\epsfbox{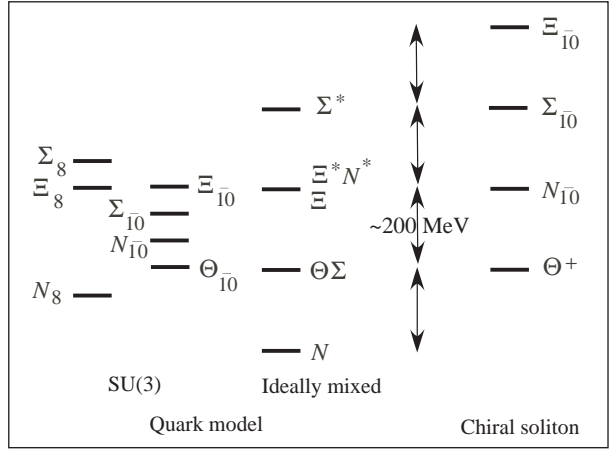}}
\centering
\caption{Expected level structure of the pentaquark 
baryons.  The most right pattern is one from the original 
chiral soliton model.  
Absolute values of the quark model values are 
set such that $\Theta^+$ appears at the same place as 
in the chiral soliton model.}
\label{splitting}
\end{figure}

This pattern of mass splitting is shown on the left 
column of Fig.~\ref{splitting}.  
The amount of the total mass difference $M(\Xi_{\bar{10}}) - M(\Theta^+)$
as shown there is
significantly smaller than the one originally estimated by 
Diakonov \etal \cite{Diakonov:1997mm}, the
spectrum of which is shown on the right side of 
Fig.~\ref{splitting}.  

As pointed out by Jaffe and Wilczek \cite{Jaffe:2003sg}, 
the antidecuplet 
nucleon and sigma states mix with the corresponding 
octet members.  
We will consider the mixing effects in more detail in 
subsection 3.4.  
Here to illustrate this mixing effect in a simple case, 
we also show in Fig.~\ref{splitting}
the mass pattern of the octet and 
the ideally mixed members.  
Since the ideally mixed states are classified by the 
strange quarks, the mass splitting between neighbors 
is $\Delta$.  
The original prediction of the 
chiral soliton model is close to this in values.  
As seen from the figure, there is significant difference 
in the mass patterns in the pentaquark baryons
depending on the realization of 
the flavor SU(3) symmetry (breaking).

%-----------Figure--------------------
%-----------Figure--------------------

In general, the constituent quark model 
can not predict absolute values of masses.  
Nevertheless, if we estimate them by using typical 
values of constituent masses, 
$m_{u}, m_{d} \sim 300$ MeV and $m_{s} \sim 500$ MeV, 
we find 
$M_{\Theta^{+}} \sim 1.7$ GeV, and other masses 
in accordance with the equi-distant rule.   
The mass of $\Theta^{+}$ is larger than the observed values.  
In Fig. \ref{splitting}, however, the mass of 
$\Theta^+$ is normalized.  

%------------------------------
\subsection{Chiral solitons}
%------------------------------

This model is based on the idea of Skyrme that 
baryons are made from weakly interacting mesons, 
solitons \cite{skyrme}.
A microscopic basis of this model is the 
$1/N_{c}$-expansion of QCD \cite{'tHooft:1974jz,Witten:1979kh,Witten:1983tw}.  
The fact that there are two light flavors is also important; 
SU(2) isospin symmetry leads to a strongly correlated 
pseudoscalar pion fields under rotations in the 
coordinate space and isospin space.  

The pion field is conveniently parametrized by an 
SU(2) matrix as 
\beq
U(\vec x) = \exp(i \vec \tau \cdot \vec \pi/f_\pi) \, , 
\label{defpi}
\eeq
where $f_\pi = 93$ MeV is the pion decay constant.  
Under the strong correlation, the hedgehog configuration is 
realized as a static solution 
where the pion field points to the radial 
direction, $\vec \pi/f_\pi = \hat r F(r)$, where the spherical 
profile function $F(r)$ is determined by solving the
classical field equation of motion.   
Nontrivial 
solutions for $F(r)$ define the ground states of the 
system.  
Due to nontrivial topology in the theory, 
different solutions $F(r)$ exist as  
classified by the 
winding number which is physically identified with the 
baryon number. 
The system of one baryon number describes the single 
nucleon sector.  

The hedgehog solution is a classical configuration and 
does not correspond to a physical nucleon state.  
To make a link between them, 
we introduce the collective variables for  
isospin rotations $A(t) \in SU(2)$, 
\beq
U(t, \vec x) = A(t) U_{H}(\vec x) A(t)^{\dagger} \, , \; \; \; 
U_H(\vec x) = \exp(i \vec \tau \cdot \hat r F(r)) \, .
\eeq
To be more precise, one needs to introduce another 
rotation in coordinate space, 
$x_{i} \to R_{ij} x_{j}$.  
The spatial rotation $R$, however, is equivalent to the 
isospin rotation $A$ due to the symmetry of the 
hedgehog configuration; 
$A$ and $R$ can not be independent degrees of freedom 
upon quantization.  
Consequently, the quantization of the $A$ variable 
leads to the wave functions which are the SU(2) 
$D$-functions for free motion in the SU(2) manifold, 
$D_{t, -m}^{I}(A)$.    
Constraints then follow in the quantized states; 
spin and isospin must take the same values; 
$J = I$ and $t = I_z, m = J_z$ \cite{hosaka_book,Adkins:1983ya}.  

When this method is applied to flavor SU(3), one finds 
several interesting consequences.  
We will state some of them without proof.  
The SU(3) baryonic states are written in terms of the 
SU(3) $D$-functions
$D^{(p,q)}_{YII_{3};Y^{R}I^{R}I_{3}^{R}}
(\alpha_{1}, \ldots \alpha_{8})$, 
where the upper and lower indices label the SU(3)
states, and $\alpha_{1}, \ldots \alpha_{8}$ are 
the Euler angles for SU(3) rotations.  
A crucial observation here is that under the hedgehog 
ansatz, the right quantum numbers 
$Y^{R}I^{R}I_{3}^{R}$
are related to the 
spin and hypercharge quantum numbers.  
Furthermore, the Wess-Zumino term puts further 
constraints on the right hyper charge and the baryon number, 
\beq
(I^{R}, I^{R}_{3}) = (J, J_{3})\, , \; \; \; 
Y^{R} =B \, , 
\eeq
where the second equation holds when $N_{c} = 3$.  
From these, it follows that the number of states of 
$Y = 1$ is $2J+1$.  
For $\bar {10}$, the state of $Y = 1$ is the nucleon and so 
$2J + 1 = 2$, or $J = 1/2$. 
The parity of this state is the same as that of the nucleon, and
hence the spin and parity of $\Theta^{+}$ and its partner 
are $J^{P} = 1/2^{+}$.  
The mass splitting among the multiplet $\bar{10}$ 
is once again equi-distant.  
In the original paper by Diakonov 
\etal ~\cite{Diakonov:1997mm}, they determined 
parameters in the mass formula
(see Eq. (\ref{mass22}) below) from information of 
non-exotic sectors, making then prediction for the exotic baryons.  
The relatively low mass of $\Theta^{+}$ was predicted this way prior 
to the observation.  
In their original work, the nucleon resonance 
$N(1710)$ was identified with a member of $\bar {10}$, 
to determine the equi-distant parameter 
$\Delta \sim$ 180 MeV.  
This pattern of the mass splitting is shown on the right 
side of Fig. \ref{splitting}.  

%%%%%%%%%%%%%%%%%

%------------------------------
\subsection{Role of chiral symmetry}
%------------------------------

It is instructive to make an interpretation of  
the results of the chiral soliton model, especially 
the fact that the $1/2^+$ state appears as the lowest 
state of $\Theta^+$.  
As it turns out, 
the role of chiral symmetry is important.  
As we have remarked in the previous subsection, 
a positive parity $\Theta^+$ requires an orbital excitation 
of a quark to an 
odd parity orbit, say $p$-orbit ($l=1$).  
This costs at least 
another $\hbar \omega = 500$ MeV for the mass of $\Theta^+$.  
A question is then whether 
there is a mechanism to lower the higher state than the 
negative parity state of $(0s)^5$.  
As shown in Ref. \cite{stancu} 
the flavor dependent 
force due to the Nambu-Goldstone boson exchanges between quarks 
has a large attraction in the pentaquark state, 
which compensates the excess of the $p$-state 
energy.

\begin{figure}[tbp] 
\epsfxsize=5cm
\centerline{\epsfbox{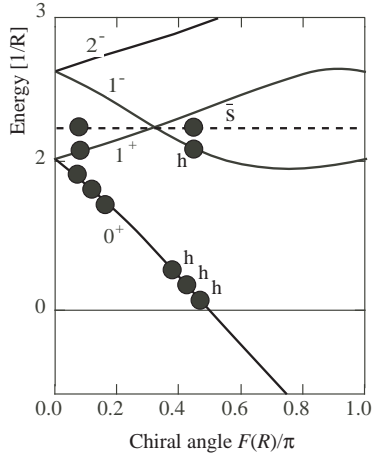}}
\caption{Quark energy levels of the chiral bag model 
as functions of the chiral angle $F$. 
For $u, d$ hedgehog quarks, energy levels vary 
as functions of $F(R)$ as denoted by $h$, while
the $s$ quark is not affected by the interaction.  
The blobs indicate how five quarks occupy the levels. }
\label{cbag_level}
\end{figure}

Here we illustrate it in the chiral bag model
by considering the quark single particle states in a
bag as functions of the chiral angle at the bag surface $F(R)$
(see Fig.~\ref{cbag_level}) \cite{Hosaka:2003jv,cbag}. 
In the presence of the pion field which 
interact with the quarks at the bag surface, the equation of 
motion for the quark field is written as
\beq
\left( i \dslash -\frac{1}{2} 
\exp(i \vec \tau \cdot \vec \pi(x)/f_\pi\, \gamma_5) 
\delta (r-R) \right) \psi = 0 \, , 
\eeq
where 
the surface $\delta$-function $\delta (r-R)$ 
indicates that the interaction occurs at the bag surface.  
In the hedgehog configuration, 
$
\vec \pi(x)/f_\pi = \hat r F(r)
$, 
the quark eigenstates are specified by the parity $P$ and 
the grand spin
which is the sum of the orbital angular momentum, spin and 
isospin, $\vec K = \vec L + \vec S + \vec I$.  
Then, the pion-quark interaction reduces to 
a spin-isospin interaction of the type
$\vec \sigma \cdot \vec \tau$.  

For a given $J = L + S = L \pm 1/2$, two $K$ values are possible, 
$K = J \pm 1/2$.  
They are degenerate when the pion-quark interaction is zero in the 
large $R$ limit as in the MIT bag model.  
As the bag radius is reduced and the pion-quark interaction 
is increased, the degeneracy is resolved. 
This phenomena is similar to the spin-orbit splitting.  
In the present case, the state of smaller $K$ is lowered, 
while the other pushed up. 
The change in the eigenenergies causes 
level crossing or a rearrangement of the pentaquark state
at a certain strength of the pion-quark interaction.
A crucial point is that the rearrangement occurs by the 
crossing of two states of opposite parities, which is 
followed by a flip of the parity of the pentaquark state.  
As shown in Fig.~\ref{cbag_level}, for small pion-quark
interaction (small $F(R)$), the five quark configuration (denoted by
the hedgehog quantum numbers $K^P$) is 
$(0^+)^4 1^+ \sim (0s)^5$, while it is replaced by the 
$(0^+)^4 1^- \sim (0s)^4 1p$  
for $F(R) \gsim 0.3 \pi$.  
In the latter, the positive parity state becomes  
the lowest for the pentaquark state.  
The parity flip occurs when the pion field is sufficiently strong.  
As is for the nucleon,  if the bag radius of $\Theta^+$ 
takes a value around $R \sim 0.6$ where the chiral angle 
$F(R) \sim \pi/2$ \cite{cbag}, 
the positive parity $\Theta^+$ can be realized.  

%-----------Figure--------------------
%-----------Figure--------------------

%------------------------------
\subsection{Model independent analysis of SU(3)}
%------------------------------

So far, we have discussed models of QCD.  
Instead, we can work out to a great extent 
by using only flavor SU(3) symmetry, and derive various 
relations among masses and coupling 
constants \cite{Pakvasa:2004pg,Hyodo:2005wa}.  
Only assumption is that particles of definite 
spin and parity belong to certain multiplets of SU(3).  
Symmetry puts constraints on mass and interaction 
hamiltonians with several parameters, which 
are determined from experimental data.  
Since there are more physical quantities than parameters, 
we can make predictions.  
If the symmetry is only approximate, 
we can estimate the breaking effect either by 
perturbation, or by preparing a wider model space 
and performing diagonalization.  

Our interest here is to clarify the nature of $\Theta^+$ and 
its partners.  
If SU(3) symmetry is good, they belong to 
pure $\bar{10}$, while if the breaking occurs (from the mass of 
strange quark), the nucleon and sigma states of the $\bar{10}$
start to mix with those of octet members.  
Presumably, we can imagine that they are also pentaquarks, 
which however is not a necessary condition \cite{Guzey:2005mc}.  
The $\Xi$ states do not mix because of the isospin symmetry.  
Therefore, we have the following set of particles to consider:
$
\Theta^+\,\;  (N_8, N_{\bar{10}}), \; 
(\Sigma_8, \Sigma_{\bar{10}}), \; 
\Xi \, .
$

We start with writing down the mass matrix 
for the antidecuplet $\bar{10}$ and octet 8,   
\beq
H =
\left(
\begin{array}{c c}
M_{\bar {10}} - aY & \delta \\
\delta & 
M_8 - bY + c [ I(I+1) - Y^2/4]
\end{array}
\right) \, , 
\label{mass22}
\eeq
where the parameters are $M_8, M_{\bar {10}}, a, b, c$ and 
$\delta$, while $Y$ and $I$ denote hyper charge and isospin.   
For $\Theta^+$ and $\Xi$ states, only the 
11 element is relevant, but for 
$N$ and $\Sigma$ states, the full $2\times 2$ matrix 
must be considered.  
The six parameters are determined by six inputs from data.  

Since spin and parity are independent of flavor, 
we can play with different $J^P$'s, and see how the 
fitting works.  
We have performed such fittings for $J^P = 1/2^-, 1/2^+ $ and 
$3/2^-$, where there are sufficient number of data.
For masses, the three choices of $J^P$ work well to a similar 
extent.  

The situation, however, changes if the  
method is applied to decay properties.  
Since the final state meson and baryon belong to 
the octet, there are two couplings from 
$\bar {10}$ and $8$.  
Fortunately these two couplings are determined 
from the decay properties of two known nucleon resonances
for the above three cases of $J^P$.  
Using the antidecuplet piece of the coupling constants, 
we can predict the decay width of $\Theta^+$.   
The results are summarized in Table \ref{thetadecay}, 
where the case of $1/2^-$ is excluded, since it gives 
too wide widths.   
Due to ambiguity in the phase of the coupling constants, 
we have two solutions as listed in the table. 
From this, we can see that the narrow decay width can be
obtained in one solution of $J^P = 3/2^-$.  

This result is natural, since for $3/2^-$, the final 
$KN$ state is d-wave, where the centrifugal 
barrier suppresses the decay amplitude.  
The narrow width of $\Theta^+$ could be due to the 
higher partial wave nature of the decaying channel.

\begin{table}[tbp]
    \centering
    \caption{Decay width of $\Theta^+$ determined from 
    the nucleon decays and the mixing angle obtained from experimental 
	masses.  
    Phase 1 corresponds to the same signs of $g_{N_8}$
    and $g_{\bar {10}}$,
    while phase 2 the opposite signs.
    All values are listed in MeV.}
    \begin{tabular}{ccll}
	$J^P$ & $\theta_N$ & Phase 1
	& Phase 2  \\
	\hline
	$1/2^+$ & $29^{\circ}$ (Mass) & 29.1 & 103.3  \\
	& $35.2^{\circ}$ (Ideal) & 49.3 & 131.8  \\
	$3/2^-$ & $33^{\circ}$ (Mass) & \phantom{0}3.1 & \phantom{0}20.0 \\
	 & $35.2^{\circ}$ (Ideal) & \phantom{0}3.9 &
	\phantom{0}21.3 \\
    \end{tabular}
    \label{thetadecay}
\end{table}

%------------------------------
\subsection{Lattice QCD}
%------------------------------

The investigation of the lattice QCD was started  
from the early stage of the development 
\cite{Csikor:2003ng,Sasaki:2003gi}.  
Employing a baryon interpolating field
with a suitable five quark configuration, the 
two point correlation function is studied.
The projection into 
a definite parity state must be also carried out.
As inspired by Ref. \cite{Sugiyama:2003zk}, 
this was first performed by 
Sasaki~\cite{Sasaki:2003gi}, who 
found a resonance-like signal slightly above the 
$KN$ threshold in the $1/2^-$ state.  
Recently, the significance of the signal has been somewhat 
weakened, being said that there is no sufficient evidence to 
deny resonances in the negative parity sector.  

By now there are several groups who performed 
simulations, but their results do not always 
agree with each other and the issue is still 
controversial \cite{Chiu:2004gg,Mathur:2004jr,Ishii:2004qe}.  
In drawing conclusions, one must know the limitation due to 
the approximations such as quenched approximation or 
finite quark mass.  
Instability of the results depending on the calculation 
scheme may indicate that the present lattice studies 
would not be accurate enough for the study of 
the pentaquark system, or that the pentaquarks might not exist.  

One of the sources of different results is the use of 
different types of interpolating field such as:
\beq
J_\Theta(x) &=& \epsilon_{abc}[u_a^T C \gamma_5 d_b]
\{ u_e (\bar s_e \gamma_5 d_c) \mp (u \leftrightarrow d) \} ,
\label{Jkn}\\
J_\Theta(x) &=& \epsilon_{abc}[u_a^T C \gamma_5 d_b]
\{ u_c (\bar s_e \gamma_5 d_e) \mp (u \leftrightarrow d) \} ,
\label{Jcsikor} \\
J_\Theta(x) &=& \epsilon_{abc}\epsilon_{aef}\epsilon_{bgh}
[u_e^TCd_f][u_g^TC\gamma_5 d_h] C \bar s_c^T .
\label{Jsasaki} 
\eeq
Ideally, if computing performance is sufficiently high, 
the result should not depend on the choice of the interpolating 
fields.  
In practice, results depend substantially on the choice.  
A possibly optimized way is to perform 
diagonalization of the results of different 
interpolating fields.  
Another problem which is physically important  
is the contamination due to the coupling to the 
non-resonant scattering state.  

Recently, an extensive analysis was performed by Takahashi \etal, 
where they considered a $2 \times 2$ matrix form of the 
correlation function generated by the two interpolating fields, 
(\ref{Jkn}) and (\ref{Jcsikor}), and the matrix was diagonalized 
to obtain states with an optimal coupling strength.  
Also, they investigated carefully the volume dependence
\cite{Takahashi:2005uk}.  
They have found a resonance like state which is rather stable 
against the change of the volume size in the $1/2^-$ sector.  
They have also studied the spectral weight factors
which also supports the resonance like nature of the 
$1/2^-$ state.  
However, the resonance signal of the $1/2^-$ channel has been
once again questioned in Ref. \cite{Holland:2005yt}.  
Recently, higher spin states have been also 
investigated \cite{Lasscock:2005kx}.  
Definitely, further study will be needed to achieve 
better understanding.  

%------------------------------
\subsection{QCD sum rule}
%------------------------------

The QCD sum rule was first applied to the pentaquarks by 
Zhu \cite{Zhu:2003ba}, and soon later by
Sugiyama \etal \cite{Sugiyama:2003zk} 
with the proper treatment 
of the parity projection.  
In this method the two point correlation function for the 
relevant baryonic state is computed in the asymptotic region in 
the operator product expansion (OPE).  
The correlation function in the asymptotic region  
is then analytically continuated to the low energy 
region to match the phenomenological spectral function.  
The method works reasonably well for the ground state 
baryons and for some resonance states, if the 
threshold parameter in the phenomenological side
is suitably chosen.  
The validity for excited states is, however, not 
well tested.  

The sum rule studies have been performed 
in many cases for the spin 1/2 sector, and signals of 
negative parity pentaquarks were seen around 1.5 GeV.  
Recently, spin 3/2 pentaquarks were also investigated 
\cite{Nishikawa:2004tk}.
The observed fact is that the OPE spectral function 
of the $1/2^+$ sector becomes negative
or unstable \cite{Sugiyama:2003zk,Kwon:2005fe}, which 
indicates either that there is no physical state 
of $1/2^+$ or that the present truncation of the OPE is not 
good enough.  
Also, there is a problem of contamination from the 
$KN$ state.  
In fact, 
Kondo \etal, claimed the importance of the 
exclusion of the $KN$ scattering state, which 
may change the result of parity~\cite{Kondo:2004cr}.  
They have performed the separation
by Fierz rearranging the operator of the type of 
(\ref{Jsasaki}) into that of $KN$ type.  
Lee \etal and Kwon \etal also estimated a $KN$ component 
in the two point function by applying the soft kaon 
theorem \cite{lee2005,Kwon:2005fe}.  
Their estimation showed only a small contribution  
of the $KN$ scattering state 
to the correlation function and therefore, 
the result of Sugiyama {\rm et. al.} is not changed. 

A fundamental question of the QCD sum rule 
is its applicability to the pentaquark sector, 
where the five-quark currents carry higher dimension 
than the ordinary baryon currents.  
In such a case, one should include 
higher orders of OPE in its asymptotic expansion.   
In this case, however, 
there emerge more operators of higher dimensions, 
the vacuum expectation values of which are not known 
well.  

%--------------------------------------------------------
\section{Decay of $\Theta^+$}
%--------------------------------------------------------

\begin{figure}[tbp] 
\epsfxsize=9cm
\centerline{\epsfbox{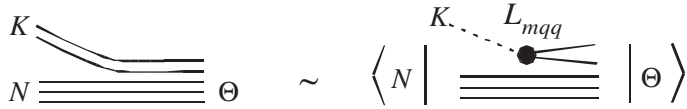}}
\caption{Decay of the pentaquark state. }
\label{fallapart}
\end{figure}

Naively, one would expect that the 
decay of the pentaquark $\Theta^+$ occurs through the 
fall-apart process as shown in 
Fig.~\ref{fallapart} (left).  
This should be so in a simple potential model for confinement.  
However, if, for instance, 
strings bind quarks, the fall-apart may not 
necessarily be relevant; if a string breaks by pulling 
two quarks apart, creation of a quark and antiquark pair must 
follow.  
Also, as discussed in Ref. \cite{Roy:2003hk}, 
the fall-apart decay was 
classified as a forbidden process; it would be interesting 
if there is such a selection rule which governs some particular  
decay processes.  
Here, we adopt the naive picture of the constituent 
quark model and test the fall-apart 
process \cite{Hosaka:2004bn}.  

%-----------Figure--------------------
%-----------Figure--------------------

For mesons and baryons of finite size, 
antisymmetrization among the four quarks is needed both for the 
initial state $\Theta^+$ and for the final $KN$ scattering states. 
In the limit of small kaon, however, the exchange term between 
quarks in the kaon and nucleon can be ignored.  
In this case, the calculation reduces to the evaluation of the 
$\Theta^+ \to N$ transition matrix 
element of the kaon source term, or equivalently of the 
axial-vector current.  
In this section we discuss the computation of this process
somewhat in detail, since such a method has not been 
explored much before in hadron physics.  

%------------------------------
\subsection{General remark}
%------------------------------

Before going to actual calculations, 
we briefly look at the general aspect for the width of 
baryons.  
Consider a decay of $\Theta^+$ going to the nucleon and 
kaon. 
Assuming the spin of the $\Theta^+$, $J=1/2$, 
the interaction lagrangian takes the form 
\beq
L_{\pm} = g_{KN\Theta} \bar \psi_N \gamma_{\pm} \psi_\Theta K \, , 
\eeq
where $\gamma_{+} = i\gamma_5$ if the parity of $\Theta^+$ is 
positive, while $\gamma_{-} = 1$ if 
the parity of $\Theta^+$ is negative.  
The decay width is then given by 
\beq
\Gamma_+ = \frac{g_{KN\Theta}^2}{2\pi} 
\frac{M_Nq^3}{E_N(E_N+M_N)M_\Theta} \, , \; \; \; 
\Gamma_- = \frac{(E_N+M_N)^2}{q^2} \Gamma_+ \, , 
\label{gamma_pm}
\eeq
for the positive ($+$) and negative ($-$) parities, 
where $M_N$ and $M_\Theta$ are the masses 
of the nucleon and $\Theta^+$, and $E_N= \sqrt{q^2+M_N^2}$ with 
$\vec q$ being the momentum of the final state kaon in the 
kaon-nucleon center of mass system, or equivalently 
in the rest frame of $\Theta^+$.  
The difference between 
$\Gamma_{\pm}$
arises due to the different coupling nature:  
$p$-wave coupling for positive parity $\Theta^+$ and 
$s$-wave coupling for negative parity $\Theta^+$, representing the 
effect of the centrifugal repulsion in the $p$-wave.   
In the kinematical point of the $\Theta^+$ decay, 
$M_\Theta = 1540$ MeV, $M_N = 940$ MeV and $m_K = 490$ MeV, 
the factor on the right hand side of (\ref{gamma_pm}) 
becomes about 50, which brings a significant difference 
in the widths of the positive and negative parity
$\Theta^+$.  
If we take $g_{KN\Theta} \sim 10$ 
as a typical strength for strong interaction coupling constants, 
we obtain $\Gamma_+ \sim 100$ MeV, while 
$\Gamma_- \sim$ 5 GeV.  
Both numbers are too large as compared with experimentally 
observed width.  
Therefore, the relevant question is whether some particular 
structure of $\Theta^+$ will suppress the above naive values, or not. 

%------------------------------
\subsection{Calculation of decay amplitudes}
%------------------------------

The matrix element of a fall-apart decay 
is written as a product of the 
spectroscopic factor and an interaction 
matrix element, 
\beq
{\cal M}_{\Theta^+ \to KN} = 
S_{KN \; {\rm in} \; \Theta^+} \cdot h_{int} \, .
\eeq
The factor
$S_{KN \; {\rm in} \; \Theta^+} \equiv
\bra (\bar s q)_K (qqq)_N | \Theta^+\ket$
is a probability amplitude of finding in the pentaquark state
three-quark and quark-antiquark clusters having the 
quantum numbers of the nucleon and kaon, 
respectively.  

Calculations of this factor was   
performed in Refs. \cite{Hosaka:2004bn,carlson}.  
It strongly depends on the internal structure of $|\Theta^+\ket$.   
Here we have investigated the four cases; 
one for the state of $(0s)^5$ of $J^P = 1/2^-$, and 
the other three for the $(0s)^4 0p$ of $J^P = 1/2^+$.  
The $(0s)^5$ configuration is unique, while 
there are four independent states for 
the $(0s)^4 0p$ configurations.  
Among them, we study the one minimizing the 
spin-flavor interaction of the type
$\sum_{i>j} (\sigma_i \sigma_j) (\lambda^f_i \lambda^f_j)$ (SF), 
the one minimizing the spin-color interaction 
$\sum_{i>j} (\sigma_i \sigma_j) (\lambda^c_i \lambda^c_j)$ (SC), 
and the one of the strong diquark correlations as proposed by 
Jaffe and Wilczek JW \cite{Jaffe:2003sg}.  
The resulting spectroscopic factors are summarized in 
Table \ref{summ_decay}.

%---------------------------------------------
\begin{table}[tbp]
\caption{\label{summ_decay} \small Spectroscopic factors and 
decay widths (in MeV) 
of $\Theta^+$ for $J^P = 1/2^{\pm}$.  
For notations SF, SC and JW, see text.  }
\begin{center}
{\small 
\begin{tabular}{c c c c c}
\hline
 & $J^P=1/2^-$ &  & $1/2^+$ & \\
 &   & SF   &   SC   &   JW   \\
\hline
$S$-factor & $1/2\sqrt{2}$ & $\sqrt{5/96}$ & $\sqrt{5/192}$ 
        & $\sqrt{5/576}$ \\ 
 $\Gamma$ (MeV) & 890 & 63 & 32 & 11 \\
\hline
\end{tabular}
}
\end{center}
\end{table}
%---------------------------------------------

Now the interaction matrix element
can be computed by the meson-quark interaction of Yukawa type:
\beq
L_{mqq} = ig \bar q \gamma_5 \lambda_a \phi^a q\, , 
\label{mqq}
\eeq
where $\lambda_a$ are SU(3) flavor matrices and 
$\phi^a$ are the octet meson fields.  
The coupling constant $g$ may be determined from the 
pion-nucleon coupling constant $g_{\pi NN} = 5g$.  
Therefore, using $g_{\pi NN} \sim 13$, we find $g \sim 2.6$.
We have calculated the $\Theta^+ \to N$ matrix elements of 
(\ref{mqq}) in the non-relativistic quark model.  
%In fact, the meson-quark interaction of (\ref{mqq}) has been 
%used for spaced like transitions, quark $\to$ quark and meson.  
%Here we need a transition of an annihilation of 
%a pair of quark and antiquark in the time like region.  
%Different kinematics changes the strength of the 
%coupling because the meson is a composite object.  
%Pentaquark decays can provide information such 
%quark and antiquark annihilation.  
%Although such a kinematical dependence may affect 
%the numerical estimation
%below, we assume to use the same coupling constant 
%as given here.  

Further details of calculation can be found in 
Ref. \cite{Hosaka:2004bn}, 
and here several results are summarized as follows.  
For the negative parity state of $(0s)^5$, the decay 
width turns out to be of order of several hundreds MeV or more, 
typically 0.5 $\sim$ 1 GeV.  
In the calculation it has been assumed that the spatial wave function 
for the initial and final state hadrons are described by a 
common harmonic oscillator states.  
Also the masses of the particles are taken as experimental 
values, e.g., $M_{\Theta^+} = 1540$ MeV.  
For the result of the negative parity state of $(0s)^5$, 
the unique prediction can be made, since there is only one 
quark model states, meaning that the state can be 
expressed in terms of the totally antisymmetrized $KN$ state.   
Since there is not a centrifugal barrier, that state
is hardly identified with a resonant state with 
a narrow width.  

For the positive parity state, 
%the orbital excitation 
%introduces more degrees of freedom for the pentaquark 
%state. 
%In fact, four independent configurations 
%are available for spin-parity $J^P = 1/2^+$ \cite{jennings}.  
%Here we consider three configurations which minimize 
%(1) a spin-flavor interaction of one meson 
%exchange \cite{carlson}, 
%(2) a spin-color interaction of one gluon 
%exchange, and 
%(3) the $S=I=0$ diquark correlated state as proposed 
%by Jaffe and Wilczek \cite{Jaffe:2003sg}.  
%The resulting decay widths are about
we obtain
$\Gamma =$ 63 MeV, 32 MeV and 11 MeV, for the 
SF, SC and JW configurations, respectively.  
The diquark correlation of (JW) develops a spin-flavor-color 
wave function having a small overlap with the decaying channel of 
the nucleon and kaon.   
In the evaluation of these values, we did not consider
spatial correlations.  
However, if, for instance, small diquarks are
developed,  spatial overlap becomes less than 
unity which further suppresses the decay width.  
In Ref. \cite{Melikhov:2004qh}, such suppression 
was shown to be significant.  
However, their absolute values of $\Gamma$ should not be 
taken seriously, since the PCAC relation was used, which 
can not be applicable to the quark model calculation.  

The small values of the decay width for $J^P = 1/2^+$ 
as compared with the large values for $J^P = 1/2^-$ can be
explained by the difference in the coupling structure; 
one is the pseudoscalar type of $\vec \sigma \cdot \vec q$
and the other the scalar type of 1.  
The former of the $p$-wave coupling 
includes a factor $q/(2M)$ which suppresses the 
decay width significantly as compared with the latter
at the present kinematics, $q \sim 250$ MeV and 
$M \sim 1$ GeV, when the same coupling constant 
$g_{NK\Theta}$ is employed.  

The present analyses can be extended straightforwardly 
to the $\Theta^+$ of spin 3/2.  
For the negative parity state, the spin 1 state of the four quarks
in the $\Theta^+$ may be combined with the spin of $\bar s$ for 
the total spin 3/2.  
In this case the final $KN$ state must be in a $d$-wave state, 
and therefore, 
the spectroscopic factor of finding a $d$-wave $KN$ state in the 
$(0s)^5$ is simply zero.  
If a tensor interaction induces an admixture of 
a $d$-wave configuration, it can decay into a $d$-wave $KN$ state.  
However, the mixture of the $d$-wave state is expected to be small
just as for the deuteron.  
There could be a possible decay channel of the nucleon and the 
vector $K^*$ of $J^P = 1^-$ \cite{Takeuchi:2004fv}.  
This decay, however, does not occur 
since the total mass of the decay channel is larger than 
the mass of $\Theta^+$.  
Hence the $J^P = 3/2^-$ state could be another 
candidate for the observed narrow state.  
This state does not have a spin-orbit partner and forms 
a single resonance peak around its energy.   
For the positive parity case, the $p$-state orbital excitation 
may be combined with the spin of $\bar s$ for the total spin 3/2. 
In this case, the calculation of the decay width is precisely 
the same as before 
(See Ref. \cite{Hosaka:2004bn} for more details).  
After taking the average over the angle $\vec q$, 
however, the coupling 
yields the same factor as for the case $J = 1/2$.  
Hence the decay rate of spin 3/2 $\Theta^+$ is the same as
that of $\Theta^+$ of spin 1/2 in the present treatment, 
if the mass of the $3/2^+$ state is the same as the 
$1/2^+$ state.

%--------------------------------------------------------
\section{Production of $\Theta^+$}
%--------------------------------------------------------

The $\Theta^+$ production from the 
non-strange initial hadrons is furnished by 
the creation of $s \bar s$ pair, which requires 
energy deposit of around 1 GeV.  
In general, the reaction mechanism of such energy 
region is not well understood.  
However, as one of practical methods, we adopt an 
effective lagrangian approach and perform computations of  
Born (tree) diagrams.   
Input parameters in the lagrangians reflect 
the properties of $\Theta^+$ and therefore, 
the comparison of calculations and experiments will help
study the structure of $\Theta^+$.  
In particular, extraction of spin and parity is the 
important purpose in the study of reactions.   
Here we briefly discuss (1) photoproduction as
originally performed in experiment by the LEPS 
group \cite{Nam:2003uf}, 
and 
(2) $\Theta^+ \Sigma^+$ production induced by 
the polarized $\vec p \vec p$ for the determination 
of the parity \cite{thomas}.  

%--------------------------------------------------------
\subsection{Photoproduction}
%--------------------------------------------------------

\begin{figure}[tbp] 
\epsfxsize=10cm
\centerline{\epsfbox{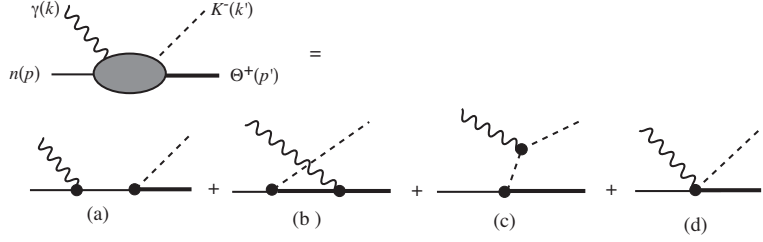}}
\caption
{Born diagrams for the $\Theta^+$ photoproduction.}
\label{photo_tree}
\end{figure}

As described in detail in Ref. \cite{Nam:2003uf}, 
in the effective lagrangian method we calculate the 
Born (tree) diagrams as depicted in Fig.~\ref{photo_tree}.  
The actual form of the interaction lagrangian depends on the 
interaction schemes, i.e., either pseudoscalar 
(PS) or pseudovector (PV).  
In the PS, the three Born diagrams (a)-(c) are 
computed with the gauge symmetry maintained.  
In the PV, on the contrary, the contact Kroll-Ruderman 
term (d) is also necessary.  
In the PS scheme, the contact term may be included in the 
antinucleon contribution of the nucleon Z-diagram.  
If chiral symmetry is respected, the low energy theorem
guarantees that the two schemes provide the same result in the 
low energy limit.   
In reality, due to the large energy deposit of order 
1 GeV, the equivalence is violated.  
It is shown that the difference in the two schemes 
is proportional to the photon momentum in the first power 
(which therefore vanishes in the low energy limit) and 
to the anomalous magnetic moment of 
$\Theta^+$ \cite{Nam:2003uf}.  

In order to express the finite size effect of the nucleon, 
we need to consider the form factor.  
Here we adopt a gauge invariant one with 
a four momentum cutoff \cite{ffgauge}.  
This form factor suppresses the nucleon pole
contributions in the PS scheme 
(and hence the contact term also in the PV scheme), 
as reflecting the fact that the nucleon intermediate state is 
far off-shell.  
Consequently, the dominant contribution is given by the 
t-channel process of the kaon exchange and/or $K^*$ meson 
exchange.  
The ambiguity of the anomalous magnetic moment of $\Theta^+$ is 
also not important.  
Therefore, the difference between the PS and PV schemes 
is significantly suppressed when kaon exchange term  
is present as for the case of the neutron target.  
This allows one to make rather unambiguous theoretical 
predictions.  

We have computed the photoproduction of $\Theta^+$ from the neutron
and proton, and first for $J^P = 1/2^{\pm}$.  
Here are several remarks:
\begin{enumerate}

\item 
When the decay width $\Gamma_{\Theta^+ \to KN} = 15$ MeV 
is used the typical total cross section values are 
about 100 [nb] for the positive parity  
and about 10 [nb] for the negative parity.  
Since the total cross section is proportional to the decay width 
$\Gamma_{\Theta^+ \to KN}$, 
experimental information on the decay width is 
important to determine the size of cross sections, 
or vise versa.  
For instance, for the decay width  about 1 MeV or less, 
the total cross sections will be of order of 10 [nb] or less 
for the positive parity and 1 [nb] or less for the negative parity.  
In general the cross sections are about ten times 
larger for the positive parity $\Theta^+$ than for the negative 
parity. 
The $p$-wave coupling $\vec \sigma \cdot \vec q$ effectively enhances the 
coupling strength by factor 3 -- 4 as compared with the $s$-wave 
coupling for the negative parity,   
when the momentum transfer amounts to 1 GeV.  

\item
For the neutron target, the kaon exchange term is dominant.  
In this case, the $K^*$ contributions are 
not important even with a large $K^*N\Theta$ coupling
$|g_{K^*N\Theta}| = \sqrt{3} |g_{KN\Theta}|$ \cite{close}.  
Hence the theoretical prediction for the neutron target 
is relatively stable.  
The angular dependence has a peak at $\theta \sim 60$ degrees 
in the center-of-mass system, a consequence of the vertex 
structure of the $\gamma KK$ vertex in the kaon exchange term.  
Since this feature is common to both parities,  
the difference in the parity of $\Theta^+$ may not be 
observed in the angular distribution.  

\item
The kaon exchange term vanishes for the case of the proton 
target.  
Therefore, the amplitude is a coherent sum of various Born terms, 
where the role of the $K^*$ exchange is also important.  
The theoretical prediction for the proton target 
is therefore rather difficult.  

\end{enumerate}

%-----------Figure--------------------
%\vspace*{0.5cm}
%-----------Figure--------------------

Very recently, the CLAS collaboration has reported no
significant evidence of $\Theta^+$ in the reaction 
$\gamma p \to \bar K^0 \Theta^+$ \cite{DeVita:2005CLAS}.  
This result should be taken seriously, 
because they have achieved significantly higher 
statistics as compared to the previous experiments
performed close to the threshold.  
What could then be the fate of $\Theta^+$?
Yet their result does not lead to the absence 
of $\Theta^+$ immediately, 
because the previous positive evidences were seen mostly in 
the reactions from the neutron.  
Due to the violation of isospin symmetry in the 
electromagnetic interaction, there could be asymmetry 
in the reactions from the proton and neutron.  
A well-known example is the Kroll-Ruderman term 
in the pion photoproduction, which survives only in the charge 
exchange channels such as $\gamma p \to \pi^+ n$.  

We have performed a calculation using once again an
effective Lagrangian, but with $J^P = 3/2^{\pm}$ 
$\Theta^+$ \cite{Nam:2005jz}.  
There is a significant difference between $1/2$ and $3/2$
cases; only PV formalism is possible for the latter case.  
Without the equivalence between the PS and PV, 
the role of the contact term in the PV scheme is very much different
for the two spin cases.  
In the $3/2$ case, the contact term dominates and the 
production rate from the neutron is large but 
that from the proton is strongly suppressed.  
Using the decay width $\Gamma \sim 1$ MeV, 
the cross section of $\gamma p \to \bar K^0 \Theta^+$
was estimated to be a few nb which does not contradict the 
CLAS data.  
Further information from the neutron target is 
crucially important to settle the problem of the 
pentaquarks.  

%--------------------------------------------------------
\subsection{Polarized proton beam and target}
%--------------------------------------------------------

This reaction was considered in order to determine the parity of 
$\Theta^+$ unambiguously, independent of any reaction 
mechanisms \cite{thomas,Nam:2004qy}.  
In the photoproduction case, the determination of parity is 
also possible if we are able to control the polarization of both 
the initial and the final states \cite{nakayama}, 
which is however very difficult in the present experimental setup.  

The system of two protons provides a selection rule due to 
Fermi statistics.  
Since the isospin is $I = 1$, the spin and angular momentum of the 
initial state must be either 
$(S, L) = (0, {\rm even})$ or $(S, L) = (1, {\rm odd})$.  
Now consider the reaction 
\beq
\vec p + \vec p \to \Theta^+ + \Sigma^+ \, .
\eeq
at the threshold region, where the relative motion in the 
final state is in $s$-wave.  
It is shown that if the initial spin state has
$S= 0$, then the parity of the final state 
is positive and hence the parity of $\Theta^+$ MUST BE
positive.  
Likewise, if $S=1$ the parity of $\Theta^+$ MUST BE
negative.   
This idea is similar to the one used to determine the parity of 
the pion \cite{panofsky}.  

One can compute production cross sections by employing an
effective lagrangian of the kaon and $K^*$ exchange model. 
The strength of the $KN\Theta^+$ vertex is determined once again 
for $\Gamma = 15$ MeV.  
The $K^*N\Theta^+$ vertex is expressed as the sum of the 
vector and tensor terms.  
The vector and tensor couplings are unknown, but here we take 
their strengths to be 
$|g^V_{K^*N\Theta}| = (1/2) |g^T_{K^*N\Theta}|
=|g_{KN\Theta}|$.  
The signs are then tested for all four possible cases
to see the effect of the couplings.  
For the $KN\Sigma$ and $K^*N\Sigma$ coupling, we take the 
phenomenological one from the Nijmegen potential \cite{stokes}.  
The monopole form factor is then introduced at each vertex, 
with the same cutoff parameter $\Lambda = 1$ GeV for simplicity.  
The choice of the interaction parameter is important for such 
high momentum transfer reactions.  
It should reflect the structure of the nucleon which has the size of 
order of 0.5 fm or larger.  
Hence $\Lambda \sim 1$ GeV is crudely the upper limit 
which is compatible 
with nucleon size $\lsim 0.5$ fm.  
In fact, the parameters of the Nijmegen soft core potential are 
chosen by such a consideration.  
 
\begin{figure}[t]
\epsfxsize=10cm
\centerline{\epsfbox{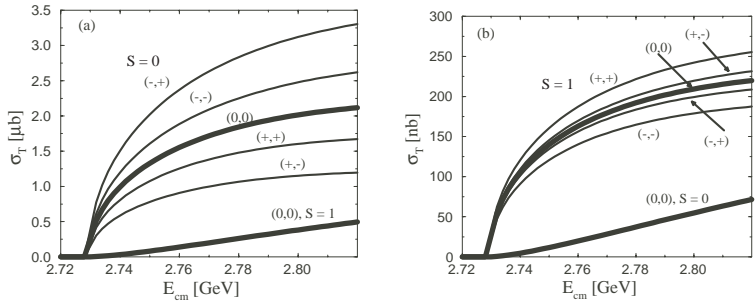}}
\caption{\small 
Total cross sections for $pp \to \Theta^+ \Sigma^+$, 
(a) for the positive parity and (b) for the negative parity, 
as functions of center of mass energy.  
Different curves correspond to different unknown 
coupling $K^*N\Theta$ \cite{Nam:2003uf}.}
\label{sigma_ppTS}
\end{figure}

The results are shown in Fig.~\ref{sigma_ppTS} for both 
positive and negative parity $\Theta^+$.
The selection rule for the positive and negative 
parity $\Theta^+$ is shown clearly by 
the energy dependence at the threshold region. 
For the allowed channel the final state is in an 
$s$-wave with the energy dependence from the threshold 
$(E - E_{th})^{1/2}$, whereas for the forbidden 
channel the partial wave of the final state is $p$-wave
with the energy dependence $(s - s_{th})^{3/2}$.  

Recently COSY-TOF reported the result for the 
$\Theta^+ \Sigma^+$ production in the unpolarized 
$pp$ scattering at $p_p = 2.95$ GeV/c \cite{cosy_tof}.  
They quote the total cross section 
$\sigma \sim 0.4 \pm 0.2$ [$\mu$b] at 30 Mev above the 
threshold in the center of mass energy.  
In comparison with theory, if we adopt a 
narrower width of about 5 MeV, the cross section 
will be about 0.5 [$\mu$b] for the positive parity 
and 0.05 [$\mu$b] for the negative parity.  
This comparison seems to favor the positive parity $\Theta^+$. 
Although there remain some ambiguities in theoretical 
calculations, such a comparison of the total cross section 
will be useful to distinguish the parity of $\Theta^+$.  

An alternative quantity which is powerful for the 
determination of the parity is the spin polarized 
quantity as defined by 
\beq
A_{xx}=(^{3}\sigma_0+^{3}\sigma_{1})/(2\sigma_{0})-1
\, , 
\eeq
where $\sigma_0$ and $\sigma_1$ are the cross sections for the 
spin singlet and triplet states of the two protons.  
By taking the ratio of the two cross sections, ambiguities of 
various coupling constants and form factors are nearly cancelled.  
The observation of $A_{xx}$ as well as the energy dependence 
of the polarized cross sections will provide the best opportunity 
to determine the parity of $\Theta^+$ 
\cite{Hanhart:2003xp,Hanhart:2004re}.

%figure---------------------------------------------
%figure---------------------------------------------

%--------------------------------------------------------
\section{Summary}
%--------------------------------------------------------

In this note we have discussed several aspects of the 
pentaquark baryon $\Theta^+$ including its structure and 
production reactions.  
Among various properties of $\Theta^+$, the importance of the 
spin and parity in relation with the decay width 
has been emphasized.  
The relevant points are as follows:
\begin{enumerate}

\item
The chiral force may change the quark energy
levels;  with a sufficient strength, an $l =1$ 
orbit may be lower than the $l=0$ orbit.
Consequently, a positive parity state 
can be the lowest pentaquark state.  
Better understanding of the role of the Nambu-Goldstone 
bosons is very important.  

\item 
The decay of the pentaquark state through the fall-apart 
process is  sensitive to the internal quark 
structure, especially to the spin and parity.  
It was shown that for $J^P = 1/2^-$, the naive ground state of 
$(0s)^5$ can no longer survive as a narrow resonance 
as the decay width is unphysically large.  
In fact, the uniqueness of the $(0s)^5$ configuration 
implies that that configuration can be written in terms of 
the $KN$ state which can no longer be bound in the naive quark 
model where the confining force vanishes between the 
two color singlet states.  
Contrary, the decay widths of $1/2^+$ states were obtained to 
be of order ten MeV, but with once again strong dependence 
on the configuration.   
We have also commented on the possibility of higher spin 
states, especially $J=3/2$.  
The $J^P = 3/2^-$ state could be an interesting alternative 
possibility.  

\item
As recently pointed by Hiyama \cite{hiyama}, 
the coupling to the $KN$ decay 
channel is extremely important when considering the 
five-body system seriously; 
it could change the nature of a confined 
pentaquark configuration completely.  
Since the five quark configuration must have components of 
two color singlet hadrons, one (or some) of them can be 
a decay channel(s), unless there are particular selection rules.  
The $J^P = 1/2^{\pm}$ states are not subject to any such selection 
rules and must be accompanied by such a coupling to the 
two hadron ($KN$) scattering state.  
In such a case, a coupled channel treatment is mandatory.  
In contrast, the $J^P = 3/2^-$ state, unless there is $d$-wave 
mixing with the $(0s)^5$ configuration, the angular momentum 
conservation forbids the coupling of the state with the 
decay channel.  

\item 
So far, our understanding of the reaction mechanism for the 
pentaquark production is rather limited.  
Perhaps, the best we can do is to use an effective 
lagrangian approach with a reasonable choice of model 
parameters.  

\item
In the photoproduction of $\Theta^+$, it was found 
a large asymmetry between the reactions from the 
proton and neutron targets, especially when 
$J = 3/2$ \cite{Nam:2005jz}.  
In relation with the understanding of the narrow 
decay width, we once again mention that 
the higher spin state would be an interesting 
possibility.  

\item
In order to determine the parity of the pentaquark, 
the polarized proton scattering 
$\vec p \vec p \to \Theta^+ \Sigma$ 
provides a model independent method.  
Measurement of such reaction is extremely important to 
further explore the physics of pentaquarks.  

\end{enumerate}

The current situation for the pentaquarks is not settled.  
Of course, the experimental confirmation is the most important 
issue. 
However, we have also seen that different theoretical 
approaches make different predictions.  
These facts imply that there could be more aspects that 
we do not know yet well about the low energy QCD.  
Perhaps the pentaquark has provided us with an ideal opportunity 
to explore further challenges to the problems.

%--------------------------------------------------------
\subsection*{Acknowledgements}
%--------------------------------------------------------

The author would like to thank 
the hospitality to the organizers of the workshop on 
HADRON PHYSICS, March 7 -- 17, (2005) Puri, India, during 
his stay.  
He thanks 
K.~Hicks, E.~Hiyama, T.~Hyodo, 
M.~Kamimura, H.C.~Kim, T.~Nakano, S.I.~Nam, M.~Oka, E.~Oset, 
A.~Titov, H.~Toki, 
A.W.~Thomas and M.J.~Vicente-Vacas for discussions and 
collaborations.  
This work supported in part by the Grant for Scientific Research
((C) No.16540252) from the Ministry of Education, Culture, 
Science and Technology, Japan.

\end{document}